\documentclass[pra,aps,twocolumn,epsfig,showpacs,superscriptaddress]{revtex4}
\usepackage{amsmath}
\usepackage{amssymb}
\usepackage[dvips]{graphicx}
\usepackage{dcolumn}
\usepackage{bm}
\usepackage[colorlinks=false,dvipdfm]{hyperref} 
\usepackage{setspace}
\usepackage{amsfonts}
\usepackage{rotating}
\usepackage{makeidx}
\usepackage{amsthm}

\newcommand{\be}{\begin{eqnarray}}
\newcommand{\ee}{\end{eqnarray}}

\begin{document}


\title{Zitterbewegung in Bogoliubov's System}

\author{Yan Li}
\affiliation{Theoretical Physics Division, Chern Institute of
Mathematics, Nankai University, Tianjin 300071, People's Republic of China}
\author{Hong-Yi Su}
\affiliation{Theoretical Physics Division, Chern Institute of
Mathematics, Nankai University, Tianjin 300071, People's Republic of China}
\author{Fu-Lin Zhang}
\email{flzhang@tju.edu.cn }
\affiliation{Physics Department, School of Science, Tianjin University, Tianjin 300072, People's Republic of China}
\author{Jing-Ling Chen}
\email{chenjl@nankai.edu.cn}
\affiliation{Theoretical Physics Division, Chern Institute of
Mathematics, Nankai University, Tianjin 300071, People's Republic of China}
 \affiliation{Centre for Quantum Technologies, National University of Singapore,
 3 Science Drive 2, Singapore 117543}

 \author{Chunfeng~Wu}
 \affiliation{Pillar of Engineering Product Development, Singapore University of Technology
and Design, 20 Dover Drive, Singapore 138682.}

\author{L.~C.~Kwek}
 \affiliation{Centre for Quantum Technologies, National University of Singapore,
 3 Science Drive 2, Singapore 117543}
 \affiliation{National Institute of Education and Institute of Advanced Studies,
 Nanyang Technological University, 1 Nanyang Walk, Singapore 637616}

\date{\today}

\begin{abstract}
We show that Bogoliubov's quasiparticle in superfluid $^3He-B$
undergoes the Zitterbewegung, as a free relativistic Dirac's
electron does. The expectation value of position, as well as spin,
of the quasiparticle is obtained and compared with that of the
Dirac's electron. In particular, the Zitterbewegung of Bogoliubov's
quasiparticle has a frequency approximately $10^5$ lower than that
of an electron, rendering a more promising experimental observation.
\end{abstract}


 \maketitle
\section{Introduction}

The phenomenon ``Zitterbewegung" (ZB), a quivering motion of a free
relativistic particle, has drawn many researchers' attention since
its theoretical prediction by Schr\"{o}dinger in 1930~\cite{Schr}.
Nonetheless, the high requirement on measuring precision has ever
since defied a direct observation. This triggered
proposals~\cite{sim1,sim2,sim3,sim4,sim5} using some other more
experimentally accessible systems to simulate this relativistic
quantum effect. For instance, a quantum simulation~\cite{sim-Nat} of
the $(1+1)$-dimensional Dirac's equation using a single trapped ion
to mimic the behavior of a free relativistic quantum particle was
performed in 2010. In this experiment, the authors studied the ZB
for different initial states and measured the particle position as a
function of time. The high-level tunability of control in this
trapped-ion experiment is also one of the merits.

The origin of ZB of a free particle (usually an electron) within the
framework of relativistic quantum mechanics could be attributed to
an interference between the positive- and negative-energy components
of wave functions. Such understanding is based on the
single-particle interpretation of the relativistic quantum
mechanics. On the other hand, in the quantum field theory which
allows particles (and antiparticles) to be temporarily created and
annihilated by satisfying Heisenberg's uncertainty principle, one
could interpret the ZB as the resulting effect of successive
scattering between an original electron and virtual
electron-positron pairs (i.e. the process of quantum vacuum
polarization)~\cite{QFT1,QFT2,QFT3}. Different relativistic quantum theories give distinct
explanations of the ZB, be that as it may, it is fairly a reasonable
perspective that the ZB should be present in a great number of
physical systems~\cite{He-A}, possibly including those in which the electron
could be in a bound state.

In this Brief Report, we investigate the ZB of a quasiparticle in
the superfluid phase of
$^3He-B$~\cite{He-B1,He-B2,He-B3,He-B4,He-B5}. Such quasiparticles
can be described by Bogoliubov's Hamiltonian, which shares the
similar expression of Dirac's Hamiltonian for a free relativistic
electron~\cite{D-to-B}\cite{Volovik}. An interesting connectiong
between Bogoliubov's Hamiltonian and Dirac's Hamiltonion has been
studied in Ref.~\cite{braid} via braiding relation. We find that the
ZB in Bogoliubov's system possesses an amplitude of order $10^{-15}
m$ and a period of order $10^{-16}s$, comparing with an amplitude of
order $10^{-12}m$ and a period of order $10^{-21}s$ for the
electron. Thus, the larger period (the lower frequency) of
Bogoliubov's quasiparticle renders a promising observation of the ZB
in $^3He-B$ comparably simpler than that of an electron.

\section{Bogoliubov's system and Zitterbewegung}

The Hamiltonian of Bogoliubov's system for quasiparticles in
$^3He-B$ can be expressed as \cite{D-to-B,Volovik,braid}
\begin{eqnarray}\label{Hamiltonian}
H_B=m(\vec{p})\beta+c\; \vec{p}\cdot\vec{\alpha},
\end{eqnarray}
where $m(\vec{p})=\vec{p}\;^2/2m-\mu$, $m$ the mass of $^3He$,
$\vec{p}$ the momentum, $\mu$ the chemical potential,
$c={\vartriangle_B}/{k_F \hbar}$, $\hbar$ the Planck constant,
$\vartriangle_B$ the equilibrium order parameter, and $k_F$ the
Fermi momentum. $\vec{\alpha}=(\alpha_1, \alpha_2, \alpha_3)$ and
$\beta$ are $4\times 4$ Hermitian matrices satisfying
\begin{subequations}
\begin{align}
&\alpha_i^2=\beta^2=1,\\
&\alpha_i\beta+\beta\alpha_i=0,\\
&\alpha_i\alpha_k+\alpha_k\alpha_i=2\delta_{ik},
\end{align}
\end{subequations} with $\delta_{ik}=1$ for $i=k$, and
$\delta_{ik}=0$ for $i\neq k$. We use the Pauli representation
\begin{eqnarray}
\vec{\alpha}=\left(\begin{array}{cc}0&\vec{\sigma}\\
\vec{\sigma}&0
\end{array}\right),\;\;\beta=\left(\begin{array}{cc}I&0\\
0&-I
\end{array}\right),
\end{eqnarray}
 where $I$ is a $2 \times 2$ unit matrix, $\vec{\sigma}=(\sigma_1, \sigma_2, \sigma_3)$ is a Pauli matrices vector with
\begin{eqnarray}
\sigma_1=\left(\begin{matrix}0&1\\
1&0
\end{matrix}\right),\;\sigma_2=\left(\begin{matrix}0&-i\\
i&0
\end{matrix}\right),\;\sigma_3=\left(\begin{matrix}1&0\\
0&-1
\end{matrix}\right).
\end{eqnarray}
 This  Hamiltonian (\ref{Hamiltonian}) becomes
``relativistic" in the limit $mc^2\gg\mu$, where it tends
asymptotically toward the Dirac Hamiltonian. However in a real $^3He
- B$, one has an opposite limit $mc^2\ll\mu$.

The wave equation of a Bogoliubov's quasiparticle can be written in
the form
\begin{eqnarray}
i\hbar \frac{\partial |\Psi(\vec{x}(t),t)\rangle}{\partial t}=H
|\Psi(\vec{x}(t),t)\rangle,
\end{eqnarray}
where $|\Psi(\vec{x}(t),t)\rangle$ is a four-component wavefunction.
The velocity of the particle is defined and calculated as
\begin{eqnarray}\label{21}
\frac{d\vec{x}(t)}{dt}&=&\frac{1}{i\hbar}\left[\vec{x}(t),H\right]
\nonumber\\
&=&\left(\frac{\vec{p}}{m}\beta(t)+c\vec{\alpha}(t)\right).
\end{eqnarray}
We should work out $\vec{\alpha}(t)$ first, and it also follows the Heisenberg equation:
\begin{eqnarray}
\frac{d\vec{\alpha}(t)}{dt}&=&\frac{1}{i\hbar}\left[\vec{\alpha}(t),H\right]\nonumber\\
&=&\frac{2}{i\hbar}(\vec{\alpha}(t)H-c\vec{p}).
\end{eqnarray}
Because the  momentum $\vec{p}$ and the Hamiltonian $H$ are
constants of motion, this equation can be integrated easily:
\begin{eqnarray}\label{alpha}
\vec{\alpha}(t)=c\vec{p}H^{-1}+\left(\vec{\alpha}(0)-c\vec{p}H^{-1}\right)e^{-2iHt/\hbar}.
\end{eqnarray}
Similarly,
\begin{eqnarray}\label{beta}
\beta(t)=m(\vec{p})H^{-1}+\left(\beta(0)-m(\vec{p})H^{-1}\right)e^{-2iHt/\hbar}.
\end{eqnarray}
Substituting (\ref{alpha}) and (\ref{beta}) into Eq. (\ref{21}), we
obtained
\begin{eqnarray}
\vec{x}(t)&=&\vec{x}(0)+\left(c^2+\frac{m(\vec{p})}{m}\right)\vec{p}H^{-1}t\nonumber\\
&&+\frac{i\hbar}{2}\left(c\vec{\alpha}(0)+\frac{\vec{p}}{m}\beta(0)-\left(c^2+\frac{m(\vec{p})}{m}\right)\vec{p}H^{-1}\right)\nonumber\\
&&\;\;\;\;\;\;\;\;\;\;\;\;\;\;\;\;\;\;\;\;\;\;\;\;\;\;\;\;\;\;\;\;\;\;\;\;\;\;\;\;\;\;\;\;\;\times\frac{e^{-2iHt/\hbar}-1}{H}.\nonumber
\end{eqnarray}


Through this part of the computation, we can see that the operators
$\vec{\alpha}$ and $\beta$ depend on time in a nontrivial way. This
adds a second term representing  a rapidly oscillating motion of the
quasiparticles over the conventional velocity operator $\vec{p}H$.
This result shows that the quasiparticle exhibits a similar
structure as a relativistic electron, and the centroid of the
wave-packet
\begin{eqnarray}
\mathcal {Z}=\biggr\langle i\hbar
\left(c\vec{\alpha}(0)+\frac{\vec{p}}{m}\beta(0)-\left(c^2+\frac{m(p)}{m}\right)\vec{p}H^{-1}\right)\nonumber\\
\times\frac{e^{-2iHt/\hbar}-1}{2H}\biggr\rangle\nonumber
\end{eqnarray}
represents a rapid oscillatory motion, i.e., the
position-\emph{Zitterbewegung},
whose amplitude is of order $ v_F\frac{\hbar}{2E}\sim10^{-15} m$ (
with $v_F=\frac{\hbar k_F}{m}$, and $E$ is the energy), and the
period is of order $\frac{\hbar}{E}\sim 10^{-16}s$, respectively.



\emph{Energy and Eigenfunction.---} Next, we need to obtain the
energy and eigenfunction of the Hamiltonian described by the
equation (\ref{Hamiltonian}). It is convenient to express the
eigenfunction in the form
 \begin{equation}\label{eif}
|\psi\rangle=\left(\begin{array}{c}u_1\\
u_2\\
u_3\\
u_4\end{array}\right)e^{i\vec{k}\cdot\vec{x}}=\left(\begin{array}{c}\phi\\
\varphi\end{array}\right)e^{i\vec{k}\cdot\vec{x}},
\end{equation}
with $\phi=\left(\begin{array}{c}u_1\\
u_2\\
\end{array}\right)\text{and} \ \varphi=\left(\begin{array}{c}
u_3\\
u_4\end{array}\right)$.
Now we define a dichotomous-valued operator
\begin{equation}\label{s}
\vec{\Sigma}
=\left(\begin{array}{cc}\vec{\sigma}&0\\
0&\vec{\sigma}
\end{array}\right),
\end{equation}
so that
\begin{equation}\label{s}
\vec{S}=\frac{\hbar}{2}\vec{\Sigma}
\end{equation}
is the spin momentum operator. It is easily to see the set of
observables $\{H, \vec{p}, \vec{\Sigma}\cdot\vec{p}\}$ commute with
one another. Since
%
%
%
\begin{equation}\label{42}
\vec{\Sigma}\cdot\vec{p}\;|\psi\rangle=\xi|\psi\rangle=\pm\hbar
k|\psi\rangle,
\end{equation}
we have
\begin{eqnarray}\label{44}
\hbar\vec{\sigma}\cdot\vec{k}\phi=\xi\phi,\\
\hbar\vec{\sigma}\cdot\vec{k}\varphi=\xi\varphi.\nonumber
\end{eqnarray}
Here $\phi$ and $\varphi$ are differed by a constant coefficient, so
we focus on $\phi$. We have 
\begin{eqnarray}
\left(\begin{array}{cc}k_3-\frac{\xi}{\hbar}&k_1-ik_2\\
k_1+ik_2&-k_3-\frac{\xi}{\hbar}
\end{array}\right)\left(\begin{array}{c}u_1\\
u_2
\end{array}\right)&=&0,
\end{eqnarray}
It is easy to get the solutions:
\begin{eqnarray}
&&{\rm for\;}\lambda=+\hbar k,\ \ \ \ \frac{u_1}{u_2}=\frac{k_3+k}{k_1+ik_2}=-\frac{k_1-ik_2}{k_3-k},\\
&&{\rm for\;}\lambda=-\hbar k,\ \ \ \
\frac{u_1}{u_2}=\frac{k_3-k}{k_1+ik_2}=-\frac{k_1-ik_2}{k_3+k},
\end{eqnarray}
with $k=|\vec{k}|$.

On the other hand, we substitute (\ref{eif}) into Bogoliubov's
Hamiltonian $H$, (i.e. $H\psi=E\psi$),
\begin{eqnarray}
\left(\begin{array}{cc}m(p)-E&c\vec{\sigma}\cdot\vec{p}\\
c\vec{\sigma}\cdot\vec{p}&-m(p)-E
\end{array}\right)\left(\begin{array}{c}\phi\\
\varphi\end{array}\right)e^{i\vec{k}\cdot\vec{r}}=0.
\end{eqnarray}
Because $\phi, \varphi$ can not all be $0$, i.e.
\begin{eqnarray}
\left|\begin{array}{cc}m(p)-E&c\vec{\sigma}\cdot\vec{p}\\
c\vec{\sigma}\cdot\vec{p}&m(p)-E
\end{array}\right|=0,\end{eqnarray}
then we obtain the spectra
\begin{eqnarray}
E=\pm\sqrt{\eta^2+c^2\hbar^2 k^2}\equiv E_{\pm},
\end{eqnarray}
with $\eta={\hbar^2 k^2}/{2m}-\mu$. 
Accordingly,
\begin{eqnarray}
{\rm for\;}E=E_+,\;\phi&=&\frac{c\hbar \vec{\sigma}\cdot\vec{k}}{\sqrt{\eta^2+c^2\hbar^2 k^2}-\eta}\varphi,\nonumber\\
{\rm for\;}E=E_-,\;\phi&=&-\frac{c\hbar
\vec{\sigma}\cdot\vec{k}}{\sqrt{\eta^2+c^2\hbar^2
k^2}+\eta}\varphi.\nonumber
\end{eqnarray}
From the equations (\ref{42}) and (\ref{44}), we know that the
eigenvalue of $\vec{\sigma}\cdot\vec{k}$ is $\pm k$. Without
explicit normalization, we list all solutions in the following:

(i) when the eigenvalues of $(H, \vec{p}, \vec{\Sigma}\cdot\vec{p})$
are $(E_+, \hbar \vec{k}, \hbar k)$, the eigenfunction is
$$|\psi_1\rangle=\frac{1}{N_1}\left(\begin{array}{c}k_1-ik_2\\
k-k_3\\
\epsilon_1{(k_1-ik_2)}\\
\epsilon_1{(k-k_3)}
\end{array}\right)e^{i\vec{k}\cdot\vec{x}},$$
where $\epsilon_1=\left(\sqrt{\eta^2+c^2\hbar^2
k^2}-\eta\right)/{c\hbar k}$.

(ii) when the eigenvalues of $\{H, \vec{p},
\vec{\Sigma}\cdot\vec{p}\}$ are $\{E_+, \hbar \vec{k}, -\hbar k\}$,
the eigenfunction is $$|\psi_2\rangle=\frac{1}{N_2}\left(\begin{array}{c}k_1-ik_2\\
-k-k_3\\
-\epsilon_2{(k_1-ik_2)}\\
\epsilon_2{(k+k_3)}
\end{array}\right)e^{i\vec{k}\cdot\vec{x}},$$
where $\epsilon_2=\epsilon_1$.

(iii) when the eigenvalues of $\{H, \vec{p},
\vec{\Sigma}\cdot\vec{p}\}$ are $\{E_-, \hbar \vec{k}, \hbar k\}$,
the eigenfunction is $$|\psi_3\rangle=\frac{1}{N_3}\left(\begin{array}{c}k_1-ik_2\\
k-k_3\\
-\epsilon_3{(k_1-ik_2)}\\
-\epsilon_3{(k-k_3)}
\end{array}\right)e^{i\vec{k}\cdot\vec{x}},$$
where $\epsilon_3=\left(\sqrt{\eta^2+c^2\hbar^2
k^2}+\eta\right)/{c\hbar k}$.

(iv) when the eigenvalues of $\{H, \vec{p},
\vec{\Sigma}\cdot\vec{p}\}$ are $\{E_-, \hbar \vec{k}, -\hbar k\}$,
the eigenfunction is $$|\psi_4\rangle=\frac{1}{N_4}\left(\begin{array}{c}k_1-ik_2\\
-k-k_3\\
\epsilon_4{(k_1-ik_2)}\\
-\epsilon_4{(k+k_3)}
\end{array}\right)e^{i\vec{k}\cdot\vec{x}},
$$
where $\epsilon_4=\epsilon_3$.
\emph{Position-Zitterbewegung.---} To study the ZB, it is necessary
to measure $\langle\vec{x}\rangle$, the expectation value of the
position operator of the quasiparticles. We introduce the energy
projection operators $\Gamma_{\pm}=\frac{1}{2}(1\pm\Lambda)$ with
$\Lambda=\frac{H}{E_p}$ and
$E_{p}=\sqrt{\left(\frac{\hbar^2k^2}{2m}-\mu\right)^2+c^2\hbar^2k^2}$.
These operators have the following properties:
\begin{eqnarray}
\Gamma_+|\psi_+\rangle&=&|\psi_+\rangle,\nonumber\\
\Gamma_-|\psi_-\rangle&=&|\psi_-\rangle,\nonumber\\
\Gamma_+|\psi_-\rangle&=&\Gamma_-|\psi_+\rangle=0
\end{eqnarray}
where $\psi_\lambda$ is the positive-energy solution and the negative-energy solution.
$\lambda=\pm1$ is the eigenvalue of the energy projection operators $\Gamma_{\pm}$.
It can be shown that
\begin{eqnarray}
[\Gamma_{\pm},
\vec{\alpha}]&=&\pm\frac{1}{2E_p}[H,\vec{\alpha}]\nonumber\\
&=&\pm\frac{1}{E_p}(c\vec{p}-\alpha H)
\end{eqnarray}
in additional,
\begin{eqnarray}
H\Gamma_\pm=\pm E_{\pm}\Gamma_\pm.
\end{eqnarray}
Since
\begin{eqnarray}
\left[\Gamma_\pm,\vec{\alpha}\right]=0,
\end{eqnarray}
after some simple calculation, we find that
\begin{eqnarray}
\Gamma_\pm \left(\vec{\alpha}(0)-c\vec{p}H^{-1}\right)\frac{e^{-2iHt/\hbar}}{2H}\Gamma_\pm=0.
\end{eqnarray}
Likewise,
\begin{eqnarray}
\Gamma_\pm \frac{\vec{p}}{m}\left(\beta-m(\vec{k})H^{-1}\right)\frac{e^{-2iHt/\hbar}}{2H}\Gamma_\pm=0.
\end{eqnarray}
Then we have the relationship
\begin{eqnarray}
\Gamma_\pm \left(c\vec{\alpha}(0)+\frac{\vec{p}}{m}\beta(0)-\left(c^2+\frac{m(\vec{p})}{m}\right)\vec{p}H^{-1}\right)\nonumber\\
\times H^{-1}e^{-2iHt/\hbar}\Gamma_\pm =0\nonumber.
\end{eqnarray}
That is to say, the oscillatory motion vanishes if the wave-packet
is a superposition of positive-energy solution only, i.e. is of the
form
\begin{eqnarray}
|\Psi\rangle=\sum\int A(\vec{p})\psi_\lambda d\vec{p}
\end{eqnarray}
with $\lambda=+1$, or of negative-energy solutions only $\lambda=-1$.
It follows that the oscillatory motion is due to interference between the positive- and negative-energy solutions
which are normally required to form a wave-packet, since neither set alone constitutes a complete set of functions.

For the superposition state $|\Psi\rangle=\sin\theta
|\psi_+\rangle+\cos\theta|\psi_-\rangle$, one may calculate the
position-Zitterbewegung as
\begin{eqnarray}
\mathcal {Z}=\frac{m(k)\hbar^2k^2-E_+^2}{E_+^2}\frac{c\hbar}{m k
E_+}\vec{k}\sin2\theta \sin\frac{2E_+t}{\hbar}
\end{eqnarray}
for $|\Psi\rangle=\sin\theta
|\psi_+\rangle+\cos\theta|\psi_-\rangle$. This is the expectation
value of oscillatory motion, that is an interference effect between
the positive and negative-energy parts. It does not appear in the
case that spinors consist entirely of positive-energy (
negative-energy) parts.


\emph{Spin-Zitterbewegung.---} Moreover, we have a look at the spin
of the Bogoliubov's system, and see whether it has the same
properties as the ZB. The spin operator satisfies the Heisenberg
equation
\begin{eqnarray}
\frac{d\vec{S}}{dt}(t)&=&\frac{1}{i\hbar}\left[\vec{S}(t),H\right]
\nonumber\\
&=&-c\vec{\alpha}(0)\times \vec{p}e^{-2iHt/\hbar},
\end{eqnarray}
which can be integrated easily, so that
\begin{eqnarray}
\vec{S}(t)
&=&\vec{S}(0)-\frac{i\hbar}{2}\left(c\vec{\alpha}(0)\times\vec{p}\right)H^{-1}\left(e^{-2iHt/\hbar}-1\right)
\end{eqnarray}
There has a rapid oscillatory motion, i.e., spin-Zitterbewegung as
\begin{eqnarray}
\mathcal {Z}_{\rm
spin}=\langle\frac{i\hbar}{2}\left(c\vec{\alpha}(0)\times\vec{p}\right)\frac{e^{-2iHt/\hbar}}{H}\rangle.
\end{eqnarray}
Similar to the above analysis, we observe
that the expectation values of spin for some types of initial
superpositions of positive- and negative-energy wavefunctions
strangely vanish. That is, $\langle \psi_i|\vec{S}|\psi_i\rangle=0,
\ \langle \psi_1|\vec{S}|\psi_2\rangle=0,\ \langle
\psi_1|\vec{S}|\psi_3\rangle=0,\ \langle
\psi_2|\vec{S}|\psi_4\rangle=0,\ \langle
\psi_1|\vec{S}|\psi_4\rangle\neq0,\ \langle
\psi_2|\vec{S}|\psi_3\rangle\neq0$.

\section{Conclusion}
To summarize, we have discussed the position-ZB in Bogoliubov's
system and shown that, besides classical uniform motion, the
centroid of the wave-packet has a rapid oscillatory motion. The
expectation value of the rapid oscillatory motion has been obtained,
indicating an interference between the positive- and negative-energy
wavefunctions
The ZB of Bogoliubov's quasiparticle has a frequency dramatically
lower than that of a free Dirac's electron, rendering a promising
observation comparably simpler. We have also discussed the spin-ZB
in Bogoliubov's system in the end.

We thank E. Solano for valuable discussion. F.L.Z. is supported by
NSF of China (Grant No. 11105097). J.L.C. is supported by National
Basic Research Program (973 Program) of China under Grant No.
2012CB921900, NSF of China (Grant Nos. 10975075 and 11175089) and
also partly supported by National Research Foundation and Ministry
of Education, Singapore.

\end{document}